\documentclass[11pt]{article}
\usepackage{amsmath,amssymb}

\numberwithin{equation}{section}	

\def\beq{\begin{equation}}
\def\eeq{\end{equation}}

\def\beqn{\begin{equation*}}
\def\eeqn{\end{equation*}}

\newcommand{\f}[2]{\frac{#1}{#2}}		
\newcommand{\s}[1]{\sqrt{#1}}			
\newcommand{\tdu}[3]{{#1_{#2}}^{#3}}

\newcommand{\Lf}[2]{\tdu{L}{#1}{#2}}		

\newcommand{\beqs}{\begin{equation}\begin{aligned}}	 
\newcommand{\eeqs}{\end{aligned}\end{equation}}

\newcommand{\beqsn}{\begin{equation*}\begin{aligned}} 
\newcommand{\eeqsn}{\end{aligned}\end{equation*}}

\newcommand{\beqc}{\begin{equation}\begin{gathered}}	
\newcommand{\eeqc}{\end{gathered}\end{equation}}

\newcommand{\beqcn}{\begin{equation*}\begin{gathered}}	
\newcommand{\eeqcn}{\end{gathered}\end{equation*}}

\newcommand{\tamphys}{\it George and Cynthia Woods Mitchell Institute
for Fundamental Physics and Astronomy,\\
Texas A\&M University, College Station, TX 77843, USA}

\newcommand{\auth}{
A. H. Mujtaba
}

\begin{document}

\begin{center}
	
	{\large{\bf Homogeneous Einstein Metrics on $SU(n)$}}
	
	\vspace{20pt}
	\auth
	
	\vspace{20pt}	
	\tamphys

	\vspace{40pt}

	\underline{ABSTRACT}

\end{center}


It is known that every compact simple Lie group admits a bi-invariant homogeneous Einstein metric. In this paper we use two ansatz to probe the existence of additional inequivalent Einstein metrics on the Lie group $SU(n)$ for arbitrary $n$. We provide an explicit construction of $(2k + 1)$ inequivalent Einstein metrics on $SU(2k)$ and $2k$ inequivalent Einstein metrics on $SU(2k+1)$.

\newpage


\section{Introduction}

The Einstein equation $R_{\mu\nu} = \lambda \, g_{\mu\nu}$, which defines an Einstein metric, constrains the Ricci and through it the Riemann curvature tensor. In general in $d$ dimensions the Ricci tensor has $\frac{1}{2}d(d+1)$ algebraically independent components while the Riemann Curvature Tensor has $\frac{1}{12}d^2(d^2-1)$ algebraically independent components \cite{nakahara}. So as $d$ increases fewer constraints are placed on the curvature of the metric. In fact for $d \geq 4$ the number of independent components of the Ricci tensor is less than that of the Riemann tensor and the gap widens as $d$ increases.

These considerations leads one to expect that as the number of dimensions $d$ increases the number of (inequivalent) Einstein metrics should also increase \cite{son_pope}. In this paper we search for some of these increasing number of Einstein metrics on $SU(n)$ group manifolds. 

The number of independent components of an $n \times n$ unitary matrix with unit-determinant are $n^2 - 1$ which means that the Lie group $SU(n)$ and its associated manifold has dimension $d = n^2 - 1$. Consequently the number of possible Einstein metrics on the $SU(n)$ group manifold increases rapidly with increasing $n$.

\subsection{Construction of metrics}

We choose to work with the vielbiens $\sigma_a$ (1-forms) coupled with a metric $g_{ab}$ (with constant components that do not vary with the parameters of the Lie-Algebra) \cite{egh}. Given a Lie group $G$ with generators $T_a$, if $g \in G$ then the left-invariant 1-forms $\sigma^a$ are given by
\beq
g^{-1} dg = \sigma^a \, T_a \, . \label{def_sigma_a}
\eeq

The general metric on the group manifold written in terms of the 1-forms is \cite{egh}
\beq
ds^2 = g_{ab} \, \sigma^a \, \sigma^b . \label{ds2}
\eeq

With the 1-forms $\sigma^a$ defined, our task is to find metrics $g_{ab}$ such that the metric $ds^2$ (as defined in (\ref{ds2})) is Einstein.

\subsection{Additional Einstein metrics}

Every simply compact Lie group admits a bi-invariant metric of the form tr$(g^{-1} dg)^2$ which in a suitable choice of basis for the generators $T_a$ can be expressed as
\beq
ds^2 = c \, \sigma^a \sigma_a ,
\eeq
where $c$ is a constant \cite{son_pope}. This corresponds to $g_{ab} = c \, \delta_{ab}$.
 
D'Atri and Ziller \cite{dazi} have shown that every simple compact Lie group, with the exception of $SU(2)$ and $SO(3)$, admits at least one additional homogeneous Einsten metric. These additional Einstein metrics, though not bi-invariant, are still invariant under the transitive $G$ action (we have chosen to preserve the full $G_L$) \cite{son_pope}. 

In particular cases, homogeneous Einstien metrics have been shown to exist in addition to the bi-invariant and D'Atri and Ziller cases. Six inequivalent homogeneous Einstein metrics have been found explicitly for the exceptional group $G_2$ \cite{gilupo}; $(3k-4)$ and $(3k-3)$ inequivalent Einstein metrics on $SO(2k)$ and $SO(2k+1)$ respectively \cite{son_pope}, and a considerable body of work in part concerning Einstein metrics on $SO(n)$, $Sp(n)$ \cite{arvan1}\cite{arvan2}\cite{arvan3}\cite{arvan4}\cite{arvan5}. These successes motivate the search for additional inequivalent Einstein metrics on $SU(n)$. 

\subsection{Inequivalence of Einstein metrics}

When searching for new Einstein metrics one must determine whether a newly found candidate is truly ``new" or if it is equivalent to an already known metric, possibly by a change of basis. A standard technique for evaluating this possibility is to calculate some dimensionless invariant quantity which is constructed from the metric and its curvature. We choose to use \cite{son_pope}\cite{gilupo} :
\beq
I_1 = R_{abcd} \, R^{abcd} \, \lambda^{-2} = |\text{Riem}|^2 \, \lambda^{-2} 
\eeq
where $R_{ab} = \lambda g_{ab}$.

For any two Einstein metrics one calculates the value of the invariant. If the calculated values are unequal then the two Einstein metrics are clearly inequivalent. If the calculated values are equal the two metrics are likely equivalent but more investigation is required to prove it conclusively.  \cite{gilupo}.


\section{Metrics on $SU(n)$}

We choose to construct and manipulate the metrics on $SU(n)$ using the left-invariant 1-forms $\Lf{A}{B}$ where $1 \leq A \leq n$. These have the property ${\Lf{A}{B}}^\dagger = \Lf{B}{A}$ and obey the algebra \cite{gilupo}
\beq \label{dL}
d\Lf{A}{B} = i \, \Lf{A}{C} \wedge \Lf{C}{B} .
\eeq

The total number of possible metrics is large given the freedom to construct the 1-forms $\sigma^a$ (from the $\Lf{A}{B}$) as well as the metric $g_{ab}$. In our search for Einstein metrics on $SU(n)$ we choose to study certain classes of ansatz. These take the form of schemes for the construction of hermitian traceless 1-forms from the $\Lf{A}{B}$ as well as particular choices for the metric $g_{ab}$.

\subsection{Scheme 1}

\subsubsection{The Generators}

The scheme consists of the construction of $n^2-1$ traceless Hermitian 1-forms $K_i$ from the $\Lf{A}{B}$. Let $m \equiv n(n-1) / 2$. We begin by constructing $n(n-1)/2$ ``traceless Hermitian" 1-forms of the form
\beq \label{def_K1}
K_i = \Lf{A}{B} + \Lf{B}{A}
\eeq
where $A \neq B$, for example $K_1 = \Lf{1}{2} + \Lf{2}{1}$. The next $n(n-1)/2$ 1-forms will be taken to be
\beq
K_{m+i} = i \, ( \Lf{A}{B} - \Lf{B}{A} )
\eeq
where $A \neq B$.

Since the 1-forms $K_i$ are obtained by taking linear combinations of the $\Lf{A}{B}$ one can describe the construction in the language of matrices. Let $\vec{l}$ be the vector with entries $l_i = \Lf{i}{i}$ and $\vec{k}$ have entries $k_i = K_{2m+i}$ for $1 \leq i \leq n$ then\footnote{Note that the last $K_i$ so defined is $K_{n^2}$ which is not a 1-form of $SU(n)$ and has non-zero trace to boot (it corresponds to the unit matrix, that is the generator of $\mathfrak{u}(1)$).} :
\beq \label{k_PQ_l}
\vec{k} = \mathbf{P} \, \mathbf{Q} \, \vec{l}
\eeq
where
\beqs \label{def_PQ}
		\mathbf{P} &= \begin{pmatrix} 
									(\f{2}{n-1} - 1) & \f{2}{n-1} & \f{2}{n-1} & \cdots & \f{2}{n-1} & 0 \\
									\f{2}{n-1} & (\f{2}{n-1} - 1) & \f{2}{n-1} & \cdots & \f{2}{n-1} & 0 \\
									\vdots & \vdots & \ddots & & \vdots & 0 \\
									\f{2}{n-1} & \f{2}{n-1} & \cdots & \f{2}{n-1} & (\f{2}{n-1} - 1) & 0 \\
									0 & 0 & 0 & \cdots & 0 & 1
					\end{pmatrix} 
	\\ 
	\\
		\mathbf{Q} &= \begin{pmatrix}
									\f{1}{\s{2}} & \f{-1}{\s{2}} & 0 & 0 & \cdots & 0 \\
									\f{1}{\s{6}} & \f{1}{\s{6}} & \f{-2}{\s{6}} & 0 & \cdots & 0 \\
									\vdots & \vdots & \ddots & \ddots & \cdots & \vdots \\
									\f{1}{\s{n(n-1)}} & \f{1}{\s{n(n-1)}} & \cdots & \cdots & \f{1}{\s{n(n-1)}} & \f{-(n-1)}{\s{n(n-1)}} \\
									\f{1}{\s{n}} & \f{1}{\s{n}} & \cdots & \cdots & \cdots & \f{1}{\s{n}}
					\end{pmatrix}
\eeqs

This construction, in particular the form of the matrix $\mathbf{P}$ was chosen to keep the $K_{2m+i}$ on a symmetric footing with respect to the $\Lf{i}{i}$. The need for symmetry arises from the consideration of the number of metric constants in our calculations. If we are able to place multiple 1-forms on a symmetric footing, in essence choosing to deal with entire subspaces as opposed to individual 1-forms, we can assign the same metric constant to them. This will reduce the number of metric constants in our calculations reducing the computational complexity of the problem.

\subsubsection{The Metric}

If we define $m = \frac{n(n-1)}{2}$ and use the definitions of $K_i$ in (\ref{def_K1}) through (\ref{def_PQ}) the metric that constitutes the ansatz for scheme 1 is given by
\beq \label{metric1}
ds^2 = x_1 \sum_{i=1}^m K_i^2 + x_2 \sum_{i=m+1}^{2m} K_i^2 + x_3 \sum_{i=2m+1}^{n^2-1} K_i^2
\eeq
With this construction the task of finding an Einstein metric is reduced to finding the values of the three constants $x_i$ for which the metric defined above is Einstein.

\subsubsection{Solutions}

We implemented the calculational algorithm using a computer program, and analyzed the results to motivate an analytical solution. The Einstein equation led to a system of three unique simultaneous equations in 4 variables, the three equations corresponding to the three classes of generators and metric constants. These equations are valid for $n \geq 2$.
\beqc
\f{n}{4} - \f{n-2}{8} \f{x_2}{x_1} + \f{1}{4} \f{x_1^2}{x_2 x_3} - \f{1}{4} \f{x_3}{x_2} -\f{1}{4} \f{x_2}{x_3} = \lambda \, x_1 \\[12pt]
\f{n+6}{16} + \f{n-2}{16} \f{x_2^2}{x_1^2} + \f{1}{4} \f{x_2^2}{x_1 x_3} - \f{1}{4} \f{x_3}{x_1} - \f{1}{4} \f{x_1}{x_3} = \lambda \, x_2 \\[12pt] 
\f{n}{8} ( 2 - \f{x_2}{x_1} - \f{x_1}{x_2} + \f{x_3^2}{x_1 x_2} ) = \lambda \, x_3 \\
\eeqc

We choose to normalize the metric constants by setting $x_2 = 1$. With this choice of normalization we have the following solutions.

The homogeneous bi-invariant metric (for $n \geq 2$):
\begin{align}
	\begin{split}
		x_1 = x_2 = x_3 = 1
	\end{split}
	\begin{split}
		\lambda = \f{n}{8} &
	\end{split}
	\begin{split}\end{split}
	\begin{split}
		\f{ \vert \text{Riem}^2 \vert }{\lambda^2} = n^2 - 1
	\end{split}
\end{align}
and the additional left-invariant metric (for $n \geq 3$):
\begin{align}
	\begin{split}
		x_1 = x_3 = \f{3n+2}{n-2}
	\end{split}
	\begin{split}
		x_2 = 1
	\end{split}
	\begin{split}
	\end{split}
	\begin{split}
		\lambda = \f{n(n-2)(5n+6)}{8(3n+2)^2}
	\end{split}
\end{align}
with
\beqn
\f{ \vert \text{Riem}^2 \vert }{\lambda^2} = \f{(2n^2 + 3n + 2)(n-1)(3n+4)}{n(5n+6)}
\eeqn

Thus we have discovered two inequivalent homogeneous Einstein metrics for each value of $n \geq 3$ using the ansatz outlined in Scheme 1.

\subsection{Scheme 2}

The ansatz in scheme 2 is based on the decomposition methodology of \cite{gilupo} and \cite{son_pope}. For $n \geq 2$ and any $0 \leq p \leq n$ we define $q \equiv n - p$. We study the decomposition
\beq
SU(p) \times SU(q) \subset SU(p+q)
\eeq

\subsubsection{The Generators} \label{s2_gen}

Once again our task, in the construction of this ansatz, is to construct 1-forms corresponding to traceless Hermitian generators from the $n^2$ 1-forms $\Lf{A}{B}$. Using the decomposition methodology we split the generators in to four classes.

\textbf{Class 1:}

From the $p^2$ $\Lf{a}{b}$ where $a,b \in \{1, 2, \dots, p\}$ we construct $(p^2 - 1)$ 1-forms according to the construction given in Scheme 1, that is we have $\f{p(p-1)}{2}$ 1-forms $(\Lf{a}{b} + \Lf{b}{a})$ for $a \neq b$, $\f{p(p-1)}{2}$ 1-forms $i (\Lf{a}{b} - \Lf{b}{a}))$ for $a \neq b$ and $(p-1)$ 1-forms constructed from a ``symmetric'' mixing of the diagonal $\Lf{a}{a}$ as in (\ref{k_PQ_l}) and (\ref{def_PQ}).

\textbf{Class 2:}

From the $q^2$ $\Lf{\alpha}{\beta}$ where $\alpha,\beta \in \{p+1, p+2, \cdots, n\}$ we construct $(q^2 - 1)$ 1-forms according to Scheme 1 and analogous to Class 1.

\textbf{Class 3:}

We construct $2 \, p \, q$ 1-forms from the off-diagonal $\Lf{a}{\beta}$ ($a \in \{1, 2, \cdots, p\}$ and $\beta \in \{p+1, p+2, \cdots, n)\}$) as follows
\begin{align}
	\begin{split}(\Lf{a}{\beta} + \Lf{\beta}{a})\end{split}
	\begin{split}i (\Lf{a}{\beta} - \Lf{\beta}{a})\end{split}
\end{align}

\textbf{Class 4:}

We end by mixing diagonal 1-forms from both $SU(p)$ and $SU(q)$. The single 1-form in this class is
\beq
q \sum_{a=1}^p \Lf{a}{a} - p \sum_{\beta=p+1}^n \Lf{\beta}{\beta}
\eeq

\subsubsection{The Metric}

The metric associated with this scheme treats each class of generators (the subspaces) as a unit and associates a single metric constant to it. 
\beq \label{metric_S2}
ds^2 = x_1 \sum_{i_1 \in \textrm{\,C1}} K_{i_1}^2 + x_2 \sum_{i_2 \in \textrm{\,C2}} K_{i_2}^2 + x_3 \sum_{i_3 \in \textrm{\,C3}} K_{i_3}^2 + x_4 \, K_{i_4} ^2
\eeq
For classes 1 and 2 this corresponds to choosing the bi-invariant metric from Scheme 1.

\subsubsection{Solutions}

We implemented the calculational algorithm using a computer program and analyzed the results to motivate an analytical solution. The Einstein equation led to a system of four unique simultaneous equations in 5 metric constants, the 4 equations corresponding to the four classes of generators and metric constants. These equations are valid for $n \geq 2$ and $p \geq 0$ with $q \equiv n - p$.
\beqc
\f{p}{8} + \f{q}{8} \, \f{x_1^2}{x_3^2} = \lambda \, x_1 \\[12pt]
\f{q}{8} + \f{p}{8} \, \f{x_2^2}{x_3^2} = \lambda \, x_2 \\[12pt]
\f{p+q}{4} - \f{(p-1)(p+1)}{8p} \, \f{x_1}{x_3} - \f{(q-1)(q+1)}{8q} \, \f{x_2}{x_3} - \f{(p+q)^2}{16} \, \f{x_4}{x_3} = \lambda \, x_3 \\[12pt]
\f{p q (p+q)^2}{16} \, \f{x_4^2}{x_3^2} = \lambda \, x_4
\eeqc
We choose to normalize the variables by setting $x_3 = 1$ which results in the following set of solutions
\begin{align} \label{s2_sol1}
x_1 = 1 && x_2 = 1 && x_4 = \f{2}{p q (p+q)} && \lambda = \f{p+q}{8} = \f{n}{8}
\end{align}

This solution for every decomposition $SU(p) \times SU(q) \subset SU(n)$ corresponds to the bi-invariant homogeneous Einstein Metric which we have already discovered in Scheme 1. The other set of solutions is
\begin{equation} \label{s2_sol2}
\begin{gathered}
\begin{aligned}x_1 = \f{pq(p+q) \pm \s{pq(p^2-1)(q^2-1)}}{q(p^2 + pq + q^2 - 1)} & \hspace{40pt} & x_2 = \f{q}{p} \, x_1\end{aligned}
\\ \\
\begin{aligned}x_4 = 2 \, \f{2p(p+q) + ((1-p^2) + (1-q^2))}{1 + pq} \, x_1 & \hspace{20pt} & \lambda = \f{q}{16 p (p+q)^2} \, x_4\end{aligned}
\end{gathered}
\end{equation}

The solutions can be divided according to the decomposition being used.

\textbf{Case 1 : $q = 0$, $p = n$}

This corresponds to Scheme 1 and gives two inequivalent metrics one of which is the bi-invariant one.

\textbf{Case 2 : $q = 1$, $p = n - 1$}

Substituting $q = 1$ in (\ref{s2_sol2}) gives us $x_1 = 1$, $x_4 = \f{2}{p(p+1)}$ and $\lambda = \f{p+1}{8}$ which corresponds exactly to the first set of solutions, equivalent to the bi-invariant metric. Thus the case $q = 1$ gives us no new inequivalent metrics on $SU(n)$.

\textbf{Case 3 : $q = p$}

Substituting $q = p$ in (\ref{s2_sol2}) makes the solution for $x_1$ degenerate and so we get one inequivalent metric rather than the usual two. Note that this case is only possible if $n$ is even.

\textbf{Case 4 :} \textit{Everything Else}

If the values of $p$ and $q$ do not correspond to any of the earlier cases we have the default situation where (\ref{s2_sol1}) results in the homogeneous bi-invariant metric and (\ref{s2_sol2}) leads to two additional inequivalent metrics. Thus this case generates \textbf{two} additional inequivalent metrics.

\section{Conclusion}

Taking in to consideration the symmetry of solutions under $p$ and $q$ exchange, the difference between even and odd $n$ and the fact that $q = 1$ generates no additional metrics; a count of the solutions from the two ansatz leads us to conclude that we can now provide an explicit construction of $(2k + 1)$ inequivalent Einstein metrics on $SU(2k)$ and $2k$ inequivalent Einstein metrics on $SU(2k+1)$. 

It is important to note that we have only tested two ans\"{a}tze and with limited complexity at that. It is likely that although we have generated many new Einstein metrics, we have not exhausted the total number of possible inequivalent Einstein metrics on $SU(n)$. 


\end{document}